\documentclass[nofootinbib,superscriptaddress,showpacs,showkeys]{revtex4}

\usepackage{graphicx,epsfig}
\usepackage{amsfonts,amsmath,amssymb,amsthm,amscd}
\usepackage{color}

\begin{document}

\title{On the design of experiments to study extreme field limits}


\author{S. S. Bulanov}
\affiliation{University of California, Berkeley, California 94720, USA}
\author{M. Chen}
\affiliation{Lawrence Berkeley National Laboratory, Berkeley, California 94720, USA}
\author{C. B. Schroeder}
\affiliation{Lawrence Berkeley National Laboratory, Berkeley, California 94720, USA}
\author{E. Esarey}
\affiliation{Lawrence Berkeley National Laboratory, Berkeley, California 94720, USA}
\author{W. P. Leemans}
\affiliation{Lawrence Berkeley National Laboratory, Berkeley, California 94720, USA}
\affiliation{University of California, Berkeley, California 94720, USA}
\author{S. V. Bulanov}
\affiliation{Kansai Photon Science Institute, JAEA, Kizugawa, Kyoto 619-0215, Japan}
\author{T. Zh. Esirkepov}
\affiliation{Kansai Photon Science Institute, JAEA, Kizugawa, Kyoto 619-0215, Japan}
\author{M. Kando}
\affiliation{Kansai Photon Science Institute, JAEA, Kizugawa, Kyoto 619-0215, Japan}
\author{J. K. Koga}
\affiliation{Kansai Photon Science Institute, JAEA, Kizugawa, Kyoto 619-0215, Japan}
\author{A. G. Zhidkov}
\affiliation{Osaka University, Osaka 565-0871, Japan}
\author{P. Chen}
\affiliation{Leung Center for Cosmology and Particle Astrophysics, National Taiwan University, Taipei 10617, Taiwan}
\author{V. D. Mur}
\affiliation{Moscow Engineering Physics Institute (State University), Moscow 115409, Russia}
\author{N. B. Narozhny}
\affiliation{Moscow Engineering Physics Institute (State University), Moscow 115409, Russia}
\author{V. S. Popov}
\affiliation{Institute of Theoretical and Experimental Physics, Moscow 117218, Russia}
\author{A. G. R. Thomas}
\affiliation{University of Michigan, Ann Arbor, Michigan 48103, USA}
\author{G. Korn}
\affiliation{ELI Beamline Facility, Institute of Physics, Czech Academy of Sciences, Prague 18221, Czech Republic}
\affiliation{Max-Planck-Institut f$\ddot{u}$r Quantenoptik, Garching 85748, Germany}

\begin{abstract}
We propose  experiments on the collision of high intensity electromagnetic pulses with electron bunches and on the collision of multiple electromagnetic pulses for studying extreme field limits in the nonlinear interaction of electromagnetic waves.  The effects of nonlinear QED will be revealed in these laser plasma experiments.    
\end{abstract}

\maketitle

\section{Introduction}

Nowadays with the fast growth of laser technology the lasers provide one of the most powerful sources of electromagnetic (EM) radiation under laboratory conditions. It was already demonstrated that the intensity of $2\times 10^{22}$ W/cm$^2$ can be achieved with the present day technology \cite{Yanovsky} and there are projects to generate pulses with the intensity up to $10^{26}$ W/cm$^2$ \cite{ELI,HiPER}. It was realized that such high intensity can be used to probe an unexplored regime of interaction of charged particles with radiation and, maybe,  the properties of the vacuum \cite{HIP2 review}. The probabilities of the processes in high intensity electromagnetic (EM) fields involving electrons, positrons, and photons depend on two parameters \cite{Ritus}:
\begin{equation}
a=\frac{eE}{m\omega c}~~~\mbox{and}~~~\chi=\frac{e\hbar\sqrt{\left( F_{\mu\nu}p_\nu\right)^2}}{m^3c^4},
\end{equation}  
where $e$ and $m$ are the charge and mass of an electron, $c$ is the speed of light, $\hbar$ is Planck constant, $\omega$ and $E$ are the EM field frequency and strength respectively, $F_{\mu\nu}$ is the tensor of the EM field, and $p$ is the electron, positron or photon momentum. The first parameter, $a$, is the dimensionless amplitude of the EM field vector-potential, and is purely classical parameter. It has a meaning of electron energy gain over a distance of one wavelength in units of its rest energy, $mc^2$. When $a$ is small the most probable are the processes with minimum possible number of photons. At $a\ll 1$ the probabilities become equal to the perturbation theory probabilities and plane waves play a role of an individual photon. When $a\sim 1$ or $a>1$ the probabilities of absorbing different number of photons become comparable and the process becomes multiphoton, \textit{i.e.}, the probability has an essentially nonlinear dependence on the field. Thus $a$ is the classical nonlinearity parameter. With the rapid development of laser technology EM pulses routinely obtain $a\gg 1$. The EM pulse with highest reported intensity had $a\approx 10^2$ \cite{Yanovsky}. On the other hand, the parameter $\chi$ has the meaning of the EM field strength in the rest frame of the particle. It is responsible for the magnitude of the quantum nonlinear effects: when $a\ge 1$ these effects are maximized for $\chi\sim 1$. All these processes in high intensity EM fields are part of a new emerging branch of physics.

In order to visualize the contribution of high intensity to different interactions of particles and fields, one can imagine a cube of theories (analogous to the cube mentioned in Ref. \cite{okun}), which is located along three orthogonal axes marked by $c$,  $\hbar$, and $a$. Then each vertex of the cube corresponds to a physical theory: $(0,0,0)$ is non-relativistic mechanics, $(c,0,0)$ is Special Relativity, and $(0,\hbar,0)$ is Quantum Mechanics. The theory that has both quantum and relativistic effects included is the Quantum Field Theory, $(c,\hbar,0)$. The classical Electrodynamics corresponds to the vertex $(c,0,a)$ and atomic, molecular and optical physics to $(0,\hbar,a)$. If the high intensity effects are included in the framework of the Quantum Field Theory, \textit{i.e.}, then the corresponding vertex $(c,\hbar,a)$ corresponds to the High Intensity Particle Physics. Thus the high intensity EM fields add a whole new dimension to the processes occurring in Quantum Field Theory, significantly changing the physics of interactions.   

\begin{figure}[tbp]
\includegraphics[width=150mm]{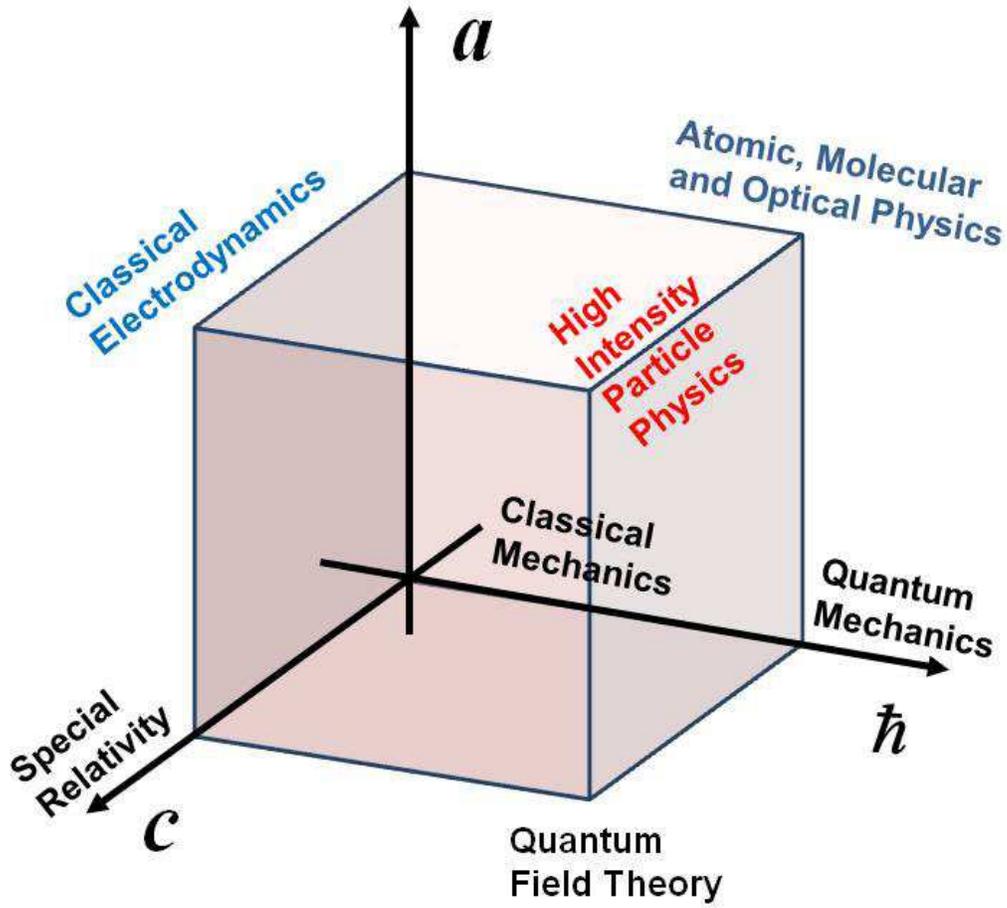}
\caption{The cube of theories.}
\end{figure}

One of the most fascinating effects of this new branch of physics is the electron-positron pair production from vacuum under the action of a strong EM field, which nonperturbative and nonlinear features can shed a light on the properties of the physical vacuum. This effect was likely first discussed by Sauter \cite{Sauter}. The vacuum-vacuum transition probability, which differs from unity in the presence of a constant uniform electromagnetic field due to $e^+e^Ð$ pair production, was found in the leading approximation by Heisenberg and Euler \cite{H&E}, and exact formulas were derived by Schwinger \cite{Schwinger}.  This effect if often referred to as the "Schwinger process". 

The probability of pair creation acquires its optimum value over the characteristic scale of the process when the electric field strength is of the order of the "critical" for quantum electrodynamics (QED) value 
\begin{equation}
E_S = m^2c^3/e\hbar =1.32\times 10^{16}~~ \mbox{V/cm}. 
\end{equation}
Such field can perform work of $mc^2$ over the Compton wavelength, $\lambda_C=\hbar/mc=3.86\times 10^{-11}$ cm. Such field strength is unattainable for static fields experimentally in the near future. Therefore attention of many researchers was focused on the theoretical study of pair creation by time-varying electric fields \cite{B&I,Popov70es,N&N,M&F,M&P,G&M&M,Ringwald,Popov00es}. However, the rapid development of laser technologies promises substantial growth of peak laser intensities. Therefore various aspects of $e^+e^-$ pair production by focused laser pulses are becoming urgent for experiments and are currently gaining much attention \cite{recent papers SSB, recent papers}. There are several other processes relevant to the study strong field effects. One of them is the emission of a photon by an electron in the EM field, usually referred to as multiphoton Thomson scattering or multiphoton Compton scattering \cite{Ritus,nCompton}. The first one being the name for the process in classical electrodynamics and the second in quantum electrodynamics, when the recoil should be taken into account. It was pointed out by Dirac that there is a possibility of transforming light into matter \cite{Dirac}, which was calculated by Breit and Wheeler \cite{Breit-Wheeler} for the collision of two photons and subsequent production of an $e^+e^-$ pair. The process of the pair production by a photon in the strong EM field was also considered in a number of papers \cite{Ritus,Reiss,laser e-beam}.    

The experimental verification of these effects, namely, the photon emission by an electron and the $e^+e^-$ pair production by a photon in the intense EM field, is limited to the E144 experiment at SLAC \cite{E144}, where a 46.5 GeV electron beam interacted with a $10^{18}$ W/cm$^2$ laser pulse. In the course of interaction, high energy photons emitted by electrons in the field of the laser pulse were able to convert to $e^+e^-$ pairs. 

\section{Principal experimental schemes}

There are two principle experimental schemes aimed at the study of particle physics effects at high laser intensity: (i) colliding laser pulses (all optical setup) and (ii) laser - e-beam interaction (See Figure 2). The first one could be employed for the study of direct vacuum breakdown, \textit{i.e.}, the $e^+e^-$ pair production under the action of intense EM field and the subsequent "avalanche". The second scheme will allow for the study of basic processes: photon emission by an electron and a photon conversion into an electron-positron pair in intense EM field. Each of the experimental setups is characterized by its own set of extreme field limits, which mark the subsequent onset of the classical radiation reaction regime, the regime when quantum recoil becomes important, and the $e^+e^-$ pair production from vacuum. 

\begin{figure}[tbp]
\epsfxsize8cm\epsffile{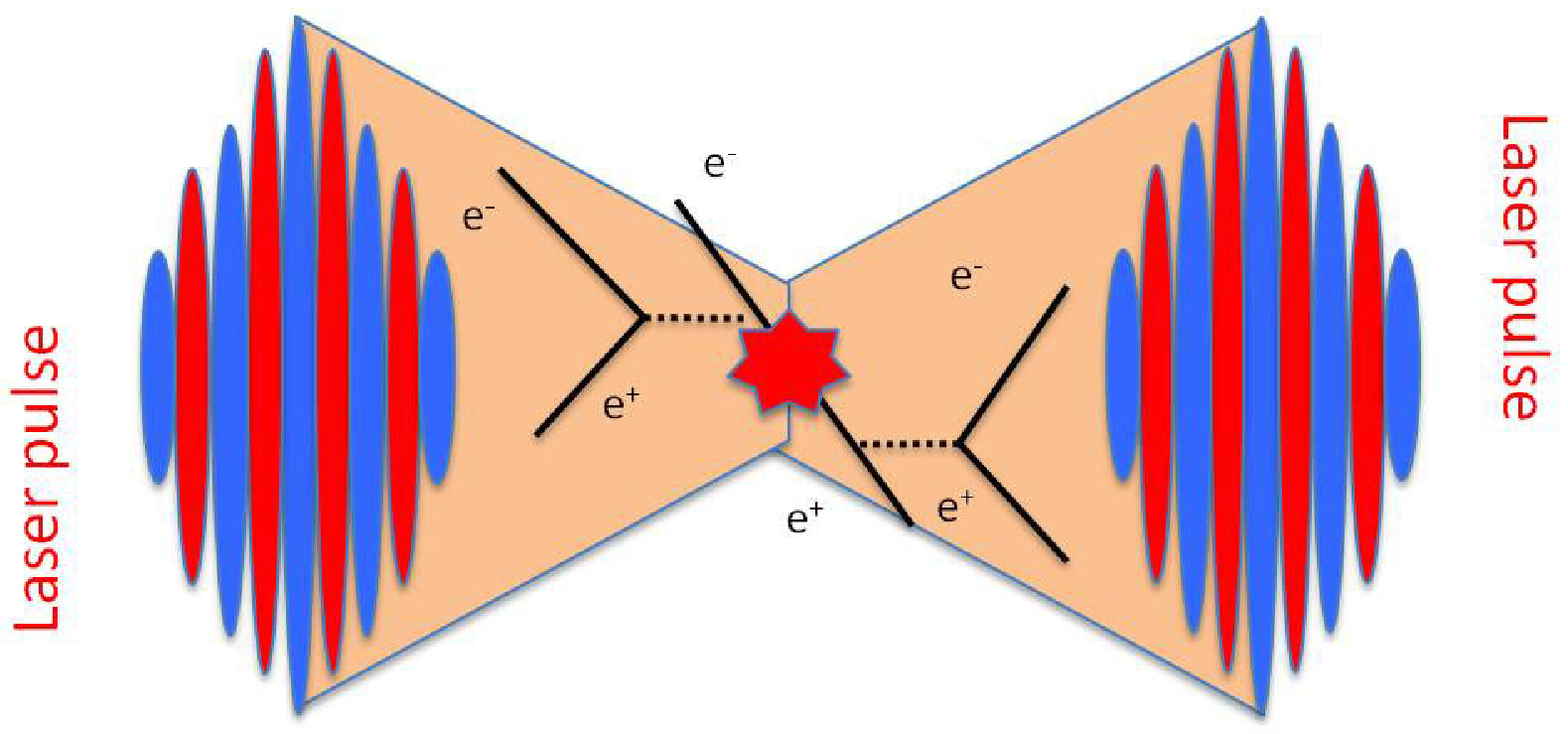} \epsfxsize8cm\epsffile{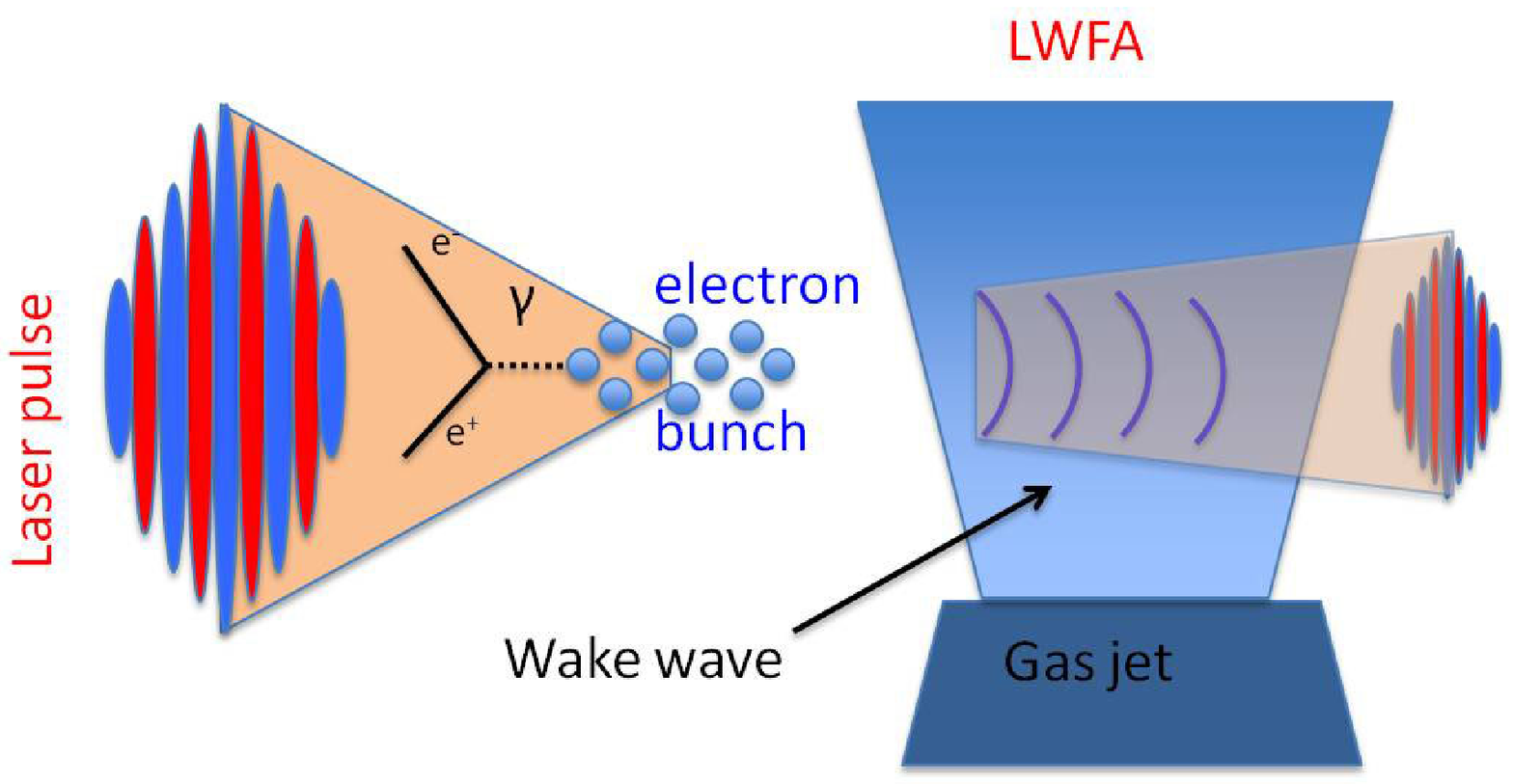} 
\caption{Principle experimental schemes aimed at the study of (i) colliding laser pulses (all optical setup); (ii) laser - e-beam interaction.}
\end{figure}

\textbf{(i) Colliding laser pulses.} Let us consider the behavior of an electron in the focus of two colliding circularly polarized laser pulses, \textit{i.e.}, in the antinode of a standing light wave with null magnetic field, and determine the thresholds for the classical radiation dominated regime of interaction, for the onset of the quantum regime, and for the Schwinger process in terms of  the laser pulse dimensionless vector-potential.  

\textbf{Classical radiation reaction.} The power emitted by the electron in the circularly polarized electric field in the ultra-relativistic limit is proportional to the forth power of its energy $P_{C,\gamma}=\epsilon_{rad}\omega m c^2\gamma_e^2(\gamma_e^2-1)\sim\gamma_e^4$, where $\epsilon_{rad}=4\pi r_e/3\lambda$ is the parameter characterizing the strength of the radiation reaction effects and $r_e=e^2/mc^2=2.8\times 10^{-13}$ cm is the classical electron radius. In the non-radiative approximation, the electron can acquire the energy from the EM field with the rate $\approx \omega mc^2 a_0$. The condition of the balance between the acquired and emitted energy is $a^3\approx\epsilon_{rad}^{-1}$. Thus the radiation reaction effects become dominant at
\begin{equation}\label{RR}
a>a_{rad}=\epsilon_{rad}^{-1/3}.
\end{equation}
For an electron moving in the focus of two 0.8 $\mu$m laser pulses the radiation reaction limit is $a_{rad}=400$ and the corresponding intensity is $I_{rad}=3.5\times 10^{23}$ W/cm$^2$. 

\textbf{Quantum regime.} The quantum effects become important when the energy of a photon emitted by an electron is of order of the electron energy, \textit{i.e.}, $\hbar\omega_m=\gamma_e mc^2$. The electron circulating in the electric field emits photons with the energy $\hbar\omega_m=\hbar\omega\gamma_e^3$. Thus the quantum recoil comes into play when 
\begin{equation}
a>a_Q=\left(\frac{2}{3}\alpha\right)^2\epsilon_{rad}^{-1},
\end{equation} 
where $\alpha=e^2/\hbar c=1/137$ is the fine structure constant. Here we took into account the fact that in the limit $I>10^{23}$ W/cm$^2$ due to strong radiation damping effects the electron energy scales as $mc^2(a/\epsilon_{rad})^{1/4}$. For an electron moving in the focus of two 0.8 $\mu$m laser pulses the quantum effects become important at $a_{Q}=1.6\times 10^3$ and the corresponding intensity is $I_Q=5.5\times 10^{24}$ W/cm$^2$.

\textbf{Schwinger limit.} As it was mentioned in the introduction the probability of pair creation acquires its optimum value over the characteristic scale of the process when the electric field strength is of the order of the "critical" for quantum electrodynamics (QED) value $E_S = m^2c^3/e\hbar =1.32\times 10^{16} ~\mbox{V/cm}$, or in terms of vector-potential:
\begin{equation}
a>a_S=\left(\frac{2}{3}\alpha\right)\epsilon_{rad}^{-1}.
\end{equation}  
For two colliding linearly polarized 0.8 $\mu$m laser pulses the "critical" field is reached at $a_S=3\times 10^5$ and the corresponding intensity is $I_S=2.3\times 10^{29}$ W/cm$^2$.
     
\textbf{(ii) Laser - e-beam interaction.} When an energetic electron bunch interacts with a counterpropagating laser pulse, the electrons of the bunch lose energy due to the radiation emission. In what follows we determine the thresholds for the classical radiation dominated regime of interaction and for the onset of quantum regime in terms of the laser pulse dimensionless vector-potential.

\textbf{Classical radiation reaction.} The motion of an electron in a EM field is governed by the equation $\dot{p}^\mu=(e/m)F^{\mu\nu}p_\nu+g^\mu$, where $g^\mu=(2e^2/3m^2)\left[\ddot{p}^\mu-(p^\mu/m c)(\dot{p}^\nu)^2\right]$ is the radiation reaction force in the Lorentz-Abraham-Dirac form \cite{LAD}. The dot over momentum stands for the differentiation over proper time $s=\int dt/\gamma_e$. If $\epsilon_{rad}a\gamma^2_e\gg 1$ then the interaction is purely dissipative and we can neglect all the EM forces except the radiation reaction. For a head-on collision of the e-beam and a laser pulse the longitudinal momentum is given by $p_x=-p_0\left[1+\epsilon_{rad}\omega (p_0/m)\int_0^t a^2(-2\eta)d\eta\right]^{-1}$. This means that the classical radiation effects become dominant at 
\begin{equation}
a>a_{rad}=\left(\epsilon_{rad}\omega\tau_{laser}\gamma_{e,0}\right)^{-1/2},
\end{equation}
where $\tau_{laser}$ is the duration of the laser pulse and $\gamma_{e,0}$ is the initial energy of the e-beam. For a 10 GeV electron beam colliding with a 0.8 $\mu$m laser pulse the radiation reaction limit is $a_{rad}=10$ and the corresponding intensity is $I_{rad}=2.2\times 10^{20}$ W/cm$^2$ for the laser pulse duration of $\omega\tau_{laser}=20\pi$. 

\textbf{Quantum regime.} The effects of quantum electrodynamics become important when the energy of an emitted photon becomes of the order of electron energy, \textit{i.e.}, $\hbar\omega_m=\gamma_e mc^2$. In this case the recoil can no longer be neglected and the probability of the $e^+e^-$ pair creation by an energetic photon in the EM field acquires its optimum value. For an electron colliding with the laser pulse a characteristic photon energy is $\hbar\omega_m\approx\hbar \omega a \gamma_e^2$ \cite{Bulanov_NIMA}, which corresponds to the condition $\chi_{e,\gamma}\sim 1$. Thus the quantum recoil comes into play when
\begin{equation}
a>a_Q=\left(\frac{2}{3}\alpha\right)\gamma_e^{-1}\epsilon_{rad}^{-1}.
\end{equation}   
For a 10 GeV electron beam colliding with a 0.8 $\mu$m laser pulse the quantum limit is $a_Q=20$ and the corresponding intensity is $I_Q=8.7\times 10^{20}$ W/cm$^2$.

\section{Probing nonlinear vacuum}

It was shown in a number of papers that the $e^+e^-$ pair production can be observed in an EM field, which is a superposition of two colliding EM pulses \cite{Popov00es, recent papers SSB}. In the antinodes of an emerging standing light wave the invariant electric field, which is responsible for pair production, acquires its maximum value. The estimates show that for tightly focussed laser pulses the intensity of $2.5\times 10^{26}$ W/cm$^2$ is needed to observe pair production, which is three orders of magnitude less than the intensity corresponding to the "critical" field strength \cite{recent papers SSB}. There are several ways to further decrease the threshold intensity, such as the utilization of multiple colliding pulses \cite{multiple} or different temporal substructures of a pulse, which may lead to quantum interference effects, inhibiting or enhancing the production probability \cite{Dunne}. However even for these additional schemes the total energy needed is around 10 kJ in 10 fs, which prevents the experimental verification of this effect in the near term.    

The problem of the avalanche formation in strong EM fields is considered in a number of recent publications \cite{avalanche} (see also \cite{Pisin}). The avalanche emerges due to the fact that if a $e^+e^-$ pair is created in the focus of two colliding  EM pulses then the created electron and positron can be very quickly accelerated by the laser field to relativistic energies and emit hard photons, which in their turn produce new $e^+e^-$ pairs. These effects have been already observed in the famous E144 SLAC experiment \cite{E144}, but yet as single events, because the energy of electrons and hard photons, as well as the laser intensity were not high enough. At high laser intensities interaction of the created electron and positron with the laser field can lead to the production of multiple new particles and thus to the formation of an avalanche-type electromagnetic discharge. 

\section{Interaction of a laser pulse with an ultra relativistic electron beam}

The scheme of the laser pulse interaction with an energetic e-beam is particularly interesting, because high energy electron beams are available from conventional accelerators and from Laser Wakefield Accelerators (LWFAs) \cite{Leemans1GeV,LWFA_review}. Moreover the simultaneous use of the laser and LWFA electrons reduces the size of the experimental setup to a table-top one. The utilization of high energy electron beams allows one to reduce the requirements to the EM field intensity to the level that can be obtained with the present day PW laser systems and still get the parameter $\chi_{e,\gamma}$ of the order of unity. Ongoing laser experiments are on the verge of observing the classical radiation reaction effects in laser matter interactions. It is believed that when the intensity of the laser as well as the energy of charged particles is high enough the interaction between them will occur in the radiation dominate regime. In this regime the charged particles will quickly lose their energy while propagating in  intense EM fields. The importance of this effect can be determined from the estimate of an energy of a particle transversing the EM field (see Eq. (\ref{RR})). For a 10 GeV e-beam interacting with a 10 cycle laser pulse the threshold value of the laser intensity is about  $10^{20}$ W/cm$^2$.

When the energy of an emitted photon is of the order of the electron energy the recoil should be taken into account. The estimates show that for $a>a_Q=(2/3)\alpha\left(\epsilon_{rad}\gamma_e\right)^{-1}$ the quantum description should be applied to such a process. This threshold value of $a$ corresponds to the EM pulse intensity of about $10^{21}$ W/cm$^2$ colliding with a 10 GeV electron beam. PW-class lasers will be able to achieve massive pair production due to high values of $a$. The process $\gamma\rightarrow e^+e^-$ acquires an optimal value at $\chi_\gamma\sim 1$ or $a_Q=(2/3)\alpha\left(\epsilon_{rad}\hbar\omega_\gamma/m c^2\right)^{-1}$, analogous to the $e\rightarrow\gamma e$ process. The observation of positron production in laser - e-beam interaction will clearly mark the threshold of High Intensity Particle Physics.  

\begin{figure}[tbp]
\epsfxsize12cm\epsffile{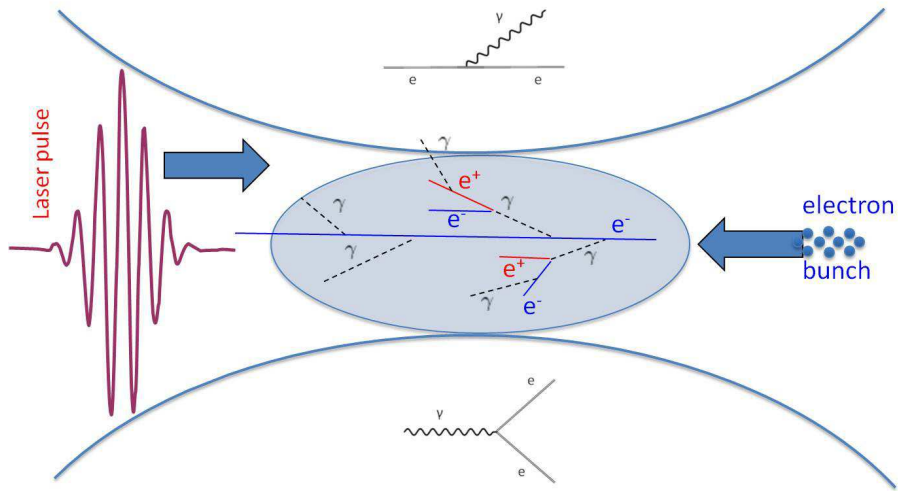}  
\caption{The scheme of the e-beam interaction with a laser pulse, and the subsequent development of a cascade-type processes.}
\end{figure}

When an energetic electron beam interacts with an intense laser pulse the electrons undergo a cascade-type process, emitting multiple photons (see Figure 3). These photons may convert into electron-positron pairs. Thus the energy of electron beam is depleted and as a result there is a massive production of electrons, positrons, and photons. The beginning of the cascade studies was marked by the E144 experiment at SLAC ($a=0.6$, $\chi_e=0.3$, and $\chi_\gamma=0.15$) \cite{E144}. The modern day laser systems can produce $a\sim 100$, and GeV-level electron beams, which not only will boost the classical and quantum nonlinearities, but also significantly reduce the characteristic scale of the process, resulting in a cascade-type multistage process in the interaction of an electron beam with an intense laser pulse. 

\begin{table}[h]
\begin{tabular}{|c|c|c|c|}
\hline  & e (150 MeV) + PW laser & e (1.25 GeV) + PW laser & e (10 GeV) + PW laser\\
\hline
$\gamma_e$ & $300$ & $2500$ & $2\times 10^4$ \\
$E/E_S$ & $3\times 10^{-4}$&  $3\times 10^{-4}$ & $3\times 10^{-4}$ \\
$\chi_e$ & $0.1$ & $0.6$ & $5$ \\
$\chi_\gamma$ & $0.01$ & $0.05$ & $1$ \\
\hline
\end{tabular}
\caption{\label{<Table>}Peak values of the invariants $\chi_e$ and $\chi_\gamma$ for a 30 fs PW laser pulse ($a\approx 100$) interacting with electron beams of different energy (150 MeV, 1.25 GeV, and 10 GeV).}
\end{table}

In Table 1 we summarize the parameters of the proposed experiment on the energetic e-beam interaction with an intense PW-class laser pulse. As it was mentioned above the interaction of a 10 GeV e-beam with an intense laser pulse is characterized by $\chi_e\sim 5$ and $\chi_\gamma\sim 1$ (the anticipated parameters on BELLA laser at then Lawrence Berkeley National Laboratory). Such experimental conditions will make it possible the study of the high intensity particle physics effects in the near term. 

\section{Conclusions}

In this paper we overview two principal schemes of experiments for the study of extreme field limits. ÄThe two colliding laser pulses scheme would allow for the study of classical radiation reaction effects ($I_{rad}\sim 3\times 10^{23}$ W/cm$^2$), the onset of the quantum regime ($I_Q\sim 5\times 10^{24}$ W/cm$^2$), and, ultimately, the $e^+e^-$ pair production from vacuum under the action of strong EM field (the so-called Schwinger process). However the high values of the intensity needed to observe these effects will prevent the realization of this scheme in the near term. In contrast, the second scheme, e-beam interaction with a laser pulse, can be realized with the present day technology. The effects of classical radiation reaction and the onset of the quantum regime can be revealed in such interaction. We showed that for a 10 GeV e-beam the laser intensity should be of the order of $10^{20}-10^{21}$ W/cm$^2$ to observe these effects. Moreover during the interaction the electrons inside the laser pulse will undergo a cascade-type multistage process involving emission of a broad spectrum of photons, and a high energy photon conversion into a $e^+e^-$ pair.   

We appreciate support from the NSF under Grant No. PHY-0935197 and the Office of Science of the US DOE under Contract No. DE-AC02-05CH11231. 

\bibliographystyle{aipproc}

\end{document}